\documentclass[aps,prl,twocolumn]{revtex4-1}
\usepackage{color}
\usepackage{ulem}
\usepackage{hyperref}
\usepackage{enumerate}

\def \be {\begin{equation}}
\def \ee {\end{equation}}
\def \ba {\begin{array}}
\def \ea {\end{array}}
\def \bea {\begin{eqnarray}}
\def \eea {\end{eqnarray}}

\def \ble {\begin{widetext}\begin{equation}}
\def \ele {\end{equation}\end{widetext}}
\def \blea {\begin{widetext}\begin{eqnarray}}
\def \elea {\end{eqnarray}\end{widetext}}

\def \and {{\textrm{and}}}

\begin{document}

\title{The area operator and fixed area states  in conformal field theories}
\author{Wu-zhong Guo}
\email{wuzhong@hust.edu.cn}
\affiliation{School of Physics, Huazhong University of Science and Technology\\
Luoyu Road 1037, Wuhan, Hubei 430074, China}

\begin{abstract}
The fixed area states are constructed by gravitational path integrals in previous studies.
 In this paper we show the dual of the fixed area states in conformal field theories (CFTs). 
 These CFT states are constructed by using spectrum decomposition of reduced density matrix $\rho_A$ for a subsystem $A$. For 2 dimensional CFTs we directly construct the bulk metric, which is consistent with the expected geometry of the fixed area states.  For arbitrary pure geometric state $|\psi\rangle$ in any dimension we also find the consistency by using the gravity dual of R\'enyi entropy. We also give the relation of parameters for the bulk and boundary state. The pure geometric state $|\psi\rangle$ can be expanded as superposition of the fixed area states. Motivated by this, we propose an area operator $\hat A^\psi$. The fixed area state is the eigenstate of $\hat A^\psi$, the associated eigenvalue is related to R\'enyi entropy of subsystem $A$ in this state. The Ryu-Takayanagi formula can be expressed as the expectation value $\langle \psi| \hat A^\psi|\psi\rangle$ divided by $4G$, where $G$ is the Newton constant. We also show the fluctuation of the area operator in the geometric state $|\psi\rangle$ is suppressed in the semiclassical limit $G\to0$. 
\end{abstract}

\maketitle


\section{Introduction}
AdS/CFT provides us a way to understand the nature of the bulk spacetime by the CFT living on the boundary\cite{Maldacena:1997re}-\cite{Witten:1998qj}. One of interesting topics in AdS/CFT is the exact duality relation between quantum states in the Hilbert space of the boundary CFT and the ones in the bulk. For some quantum states in the CFT we expect they can be effectively decribed by the classical geometries in the limit $G\to 0$.\\
The geometry is  associated with the entanglement entropy (EE) $S(\rho_A)$ of a boundary subregion $A$ by the well-known Ryu-Takayanagi formula\cite{Ryu:2006bv} for  the bulk metric with time reflection symmetry,
\bea\label{RT}
S(\rho_A)=\frac{\text{Area}(\gamma_A)}{4G},
\eea 
where $\gamma_A$ is the minimal surface in the bulk that is homology to $A$, $\rho_A$ denotes the reduced density matrix of $A$. For general bulk spacetime one should take $\gamma_A$ to be Hubeny-Rangamani-Takayanagi (HRT) surface \cite{Hubeny:2007xt}. The RT formula shows the secret relation betwen spacetime and intrinsic entanglement of underlying degree of freedoms of quantum gravity\cite{VanRaamsdonk:2010pw}.\\
The area law like relation is generalized to the holographic R\'enyi entropy by Dong \cite{Dong:2016fnf}. The R\'enyi entropy, defined as $S(\rho_A):=\frac{\log tr \rho_A^n}{1-n}$,  is  one parameter generalization of entanglement entropy. The gravity dual of R\'enyi entropy is given by 
\bea\label{holoRenyi}
n^2\partial_n \left(\frac{n-1}{n}S_n(\rho_A)\right)=\frac{\text{Area}(\mathcal{B}_n)}{4G},
\eea
where $\mathcal{B}_n$ deontes the cosmic brane with the tension $T_n=\frac{n-1}{4nG}$. The cosmic brane backreacts on the geometry by creating conical defect with opening angle $ \theta=\frac{2\pi}{n}$.\\
~\\
The R\'enyi entropies contain more information on the density matrix $\rho_A$. Actually one could construct the spectrum decompostion once knowing the R\'enyi entropy for all the index $n$. For the pure state $|\psi\rangle$ we have Schmidt decomposition $|\psi\rangle=\sum_i e^{-\frac{b^\psi+t_i}{2}}|i\rangle^\psi |\bar i\rangle^\psi$, $|i\rangle^\psi$ and $|\bar i\rangle^\psi$ are Schmidt basis of $A$ and its complementary part $\bar A$. The reduced density matrix $\rho^\psi_A$ has the spectrum decomposition $\rho^\psi_A=\sum_i e^{-b^\psi-t_i} |i \rangle \langle i|$ with $t_i \in [0,+\infty)$, where $e^{-b^\psi}$ is the  maximal eigenvalue of $\rho^\psi_A$. By definition we have $b^\psi=\lim_{n\to \infty} S^n(\rho_A^\psi)$. The modular Hamilatonian $H^\psi_A:= -\log \rho^\psi_A$ satisfies $H^\psi_A |i\rangle^\psi=(t_i+b)|i\rangle^\psi$. For quantum field theory the spectra of $H^\psi_A$ should be continuous.
By definition of R\'enyi entropy we have the relation $\sum_i e^{-n(b^\psi+t_i)} =:\int_0^\infty dt \mathcal{P}^\psi(t) e^{-n(b^\psi+t)} = e^{((1-n)S_n(\rho^\psi_A))}$, where $\mathcal{P}^\psi(t):= \sum_k \delta(t_k-t)$ is the density of eigenstates . By an inverse Laplace transformation in the variable $n$ we can obtain $\mathcal{P}^\psi(t)$\cite{Calabrese1}, see also \cite{Hung:2011nu}\cite{Guo:2020roc}. With this we can construct the state, 
\bea\label{fixareacft}
|\Phi\rangle_t^\psi:=\frac{1}{\sqrt{\mathcal{P}^\psi(t)}} \sum_k | k\rangle^\psi | \bar k\rangle^\psi \delta(t_k-t).
\eea
The reduced density matrix of $A$ is $\rho^\psi_{t,A}=\frac{1}{\mathcal{P}^\psi(t)}\sum_k |k\rangle^\psi ~^\psi\langle k|\delta(t_k-t)$. One of the interesting result is the pure geometric state $|\psi\rangle$ can be approximated by the state $\rho^\psi_{\bar t}$ with $\bar t=S(\rho^\psi_A)-S_{\infty}(\rho^\psi_A)$  for the holographic CFTs\cite{Guo:2020roc}.
Moreover, this state has flat spetra, thus the R\'enyi entropy is independent with $n$. The so-called fixed area states constructed in \cite{Dong:2018seb} also show the same property. Their approach is based on the graviational path integral with inserting a cosmic brane fixed to be on the RT surface. Roughly, we can take the condition that  R\'enyi entropies don't depend on $n$  as the definition of the fixed area state. They also has nice interpretation by the quantum error-correction code of AdS/CFT\cite{Almheiri:2014lwa}-\cite{Harlow:2016vwg}.  More discussions on the fixed area states can be found in \cite{Akers:2018fow}-\cite{Dong:2021clv}.  \\
~\\
One of the motivation of this paper is to show the state $|\Phi\rangle^\psi_t$ is exactly dual to the fixed area state for any $t \sim O(c)$. For the vacuum case in AdS$_3$ and $A$ being an interval, we show this conclusion by direct constructions of the bulk geometry and calculation of the R\'enyi entropy in the bulk. For more general cases, combination of the proposal of holographic R\'enyi entropy (\ref{holoRenyi}), we show the state $|\Phi\rangle^\psi_t$ corresponds to the geometry by inserting a cosmic brane $\mathcal{B}_{n^*}$ with tension $\mu_{n^*}=\frac{n^*-1}{4 n^*G}$, where $n^*$ can be solved by an equation invloving $S_n(\rho^\psi_A)$ and $t$.\\
With these results we conclude any pure geometric state can be taken as superposition of fixed area states. The coefficients of the superposition are associated with the area of the cosimic brane $\mathcal{B}_{n^*}$. Motivated by this, we introduce an area operator $\hat{A}$, for which the fixed area states are its eigenstates. The expectation value of $\hat A$ divided by $4G$ in the geometric state gives the EE. This can be seen as a quantum version of RT formula. 

\section{Vacuum state: an explicit example }

Consider a  2-dimensional CFT with central charge $c$ on complex plane  with the coordinate $(w,\bar w):=(x+i\tau, x-i\tau)$. In this section we will remove the superscript ``$\psi$'' to indicate the quantities are defined for vacuum state.  For an interval $A=[-R,R]$ in the vacuum state the R\'enyi entropy is universal for 2D CFTs \cite{Calabrese:2004eu}, given by $S_{n}(\rho_A)=(1+\frac{1}{n})b$ with $b:=\lim_{n\to \infty}S_{n}(\rho_A)=\frac{c}{6}\log \frac{2R}{\epsilon}$. We can obtain the density of  eigenstates with respect to $t$ \cite{Calabrese1}
\bea\label{densityvacuum}
\mathcal{P}(t)=\delta(t)+\sqrt{\frac{b}{t}} I_1(2\sqrt{b t})H(t),
\eea    
where $I_n(z)$ is the modified Bessel function of the first kind, $H(t)$ is the Heaviside step function. \\
In the CFT side the EE of the state $\rho_{t,A}$ is given by $\log \mathcal{P}(t)$. 
For the holographic CFT $b\sim O(c)\gg 1$, taking $t$ to be the order of $c$. The density of state  $\mathcal{P}(t)\simeq \frac{b e^{2\sqrt{bt}}}{\sqrt{4\pi} (bt)^{3/4}}$.
The EE is 
$\log \mathcal{P}(t)\simeq 2\sqrt{bt}+O(\log c)$. \\
By construction the R\'enyi entropy of the state $\rho_{t,A}$ is same as the EE, which is an important feature for the so-called fixed area states \cite{Dong:2018seb}. In the following we would like to show the state $|\Phi\rangle_t$ is a fixed area state by explicitly constructing the bulk geometry.\\
Using a similar method as in \cite{Kraus:2016nwo}, we can get the expectation value of  stress energy tensor $T(w)$ in the state $|\Phi\rangle_t$\cite{Guo:2020roc} 
\bea\label{onepointT}
\langle T(w)\rangle_t=\frac{cR^2}{6(R^2-w^2)^2}  \left(1-\frac{t}{b}\right).
\eea
Similarly, one could get $\langle \bar T(\bar w)\rangle_t$ by replacing $w$ with $\bar w$ in the above expression. For pure state the R\'enyi entropy satisfies $S_n(\rho_A)=S_n(\rho_{\bar A})$, where $\rho_{\bar A}$ is the reduced density matrix of the complementary part of $A$. 
The singularity at the ending points of interval $A$ is associated with the conical defect as we will show soon. 
The bulk solution is fixed by the one-point function of $T(w)$, the bulk geometry dual to $|\Phi\rangle_t$ is  
\bea\label{geometrydual}
ds^2=\frac{dy^2}{y^2}+\frac{L_t}{2}dw^2+\frac{\bar L_t}{2}d\bar w^2+\left(\frac{1}{y^2}+\frac{y^2}{4}L_t \bar L_t\right)dwd\bar w,\nonumber\\
~
\eea
where $L_t:=-\frac{12}{c}\langle T(w)\rangle_t$, $\bar L_t:=-\frac{12}{c}\langle \bar T(w)\rangle_t$. The above solution has singularity in the coordinate $(y,w,\bar w)$.  By a conformal transformation 
$\xi=\left(\frac{R+w}{R-w}\right)^\alpha, \bar \xi =\left(\frac{R+\bar w}{R-\bar w}\right)^\alpha$
with $\alpha:=\sqrt{\frac{t}{b}}$, we have $\langle T(\xi)\rangle =\langle \bar T(\bar \xi)\rangle \rangle=0$. At the points $\xi=0,\infty$ has conical defect with opening angle $\theta=2\pi \alpha $.  
The dual bulk solution is the Poincar\'e coordinate $ds^2=\frac{d\eta^2+d\xi d\bar \xi}{\eta^2}$ with a conical defect line $\gamma$.  \\
With the geometry (\ref{geometrydual}) one could find the geodesic line $\gamma_A$ connecting the ending points of $A$ and evaluate the holographic EE  by  using the RT formula (\ref{RT}).  The details of the calculations can be found in Appendix A.  The result is 
\bea\label{holoEEvacuum}
S_A(\rho_{t,A})=\frac{L_{\gamma_A}}{4G} =\frac{\alpha c}{3}\log\frac{2R}{\epsilon}=2\sqrt{bt},
\eea
where we have used the Brown-Henneaux relation $c=\frac{3}{2G}$\cite{Brown:1986nw}. The result is exactly consistent with the CFT result to the leading order in $1/G$.
 \\
Consider the $n$-replica state $ \rho_{t,A}^n$, the one-point function $tr(\rho_{t,A}^n T(w)) $ is given by the same formula as (\ref{onepointT}). But now $w$ is the coordinate on the $n$-sheet Riemann surface $\mathcal{R}_n$.  Adopting polar coordinates near the ending points of $A$, we have $w-R\simeq r e^{i\theta}$ with $\theta\sim \theta +2 n \pi$. Using the same conformal transformation $w\to \xi=\left(\frac{w+R}{w-R}\right)^\alpha$, $\mathcal{R}_n$ is mapped to the $\xi$-plane with the conical defect with opening angle $\theta_n=2\pi n \alpha$. Therefore, the dual bulk geometry $\mathcal{M}_n$ for $\mathcal{R}_n$ is the Poincar\'e coordinate with a conical defect line  $\gamma$. Moreover, $\mathcal{M}_n$ can be constructed by cyclically gluing n-copy geometry (\ref{geometrydual}) together along the defect line $\gamma$. In \cite{Dong:2018seb} the fixed area states are constructed in the way as we have stated above. The conical defect line can be realized by inserting codimension-2 cosmic branes (lines in AdS$_3$). The tension of the cosmic brane $\mu_n$ is associated with the parameter $\alpha$ by the relation $\mu_n= \frac{1-n \alpha}{4G} $\cite{Vilenkin:1981zs}.\\
We can show the defect line $\gamma=\gamma_A$ by using the requirement that $S_n(\rho_{t,A})=S(\rho_{t,A})$. To evaluate the R\'enyi entropy $S_n(\rho_{t,A})$ we need to know the bulk action $I_{bulk}(n)$ ,which includes the on shell action $I_{g}(n)$ of the geometry $\mathcal{M}_n$ and the brane action $I_b(n)$. We show the details of the calculations in Appendix B. The result is 
\bea\label{ncopyaction}
I_{bulk}(n)=I_g(n)+I_b(n),
\eea
with 
\bea
&&I_g(n)=\frac{n}{4G}(1-\alpha^2)\log\frac{2R}{\epsilon}=n(1-\alpha^2)b,\nonumber \\
&&I_b(n)=\frac{(n\alpha-1)L_\gamma}{4G},
\eea
where $L_\gamma$ is the length of the defect line $\gamma$.  The R\'enyi entropy is 
\bea
S_n(\rho_{t,A})=\frac{I_{bulk}(n)-nI_{bulk}(1)}{n-1}=\frac{L_\gamma}{4G}.
\eea
Comparing with the holographic EE result (\ref{holoEEvacuum}) we have $L_{\gamma}=L_{\gamma_A}$. This means the 
defect line $\gamma$ coincides with the geodesic line $\gamma_A$. \\
We expect the states (\ref{fixareacft}) are dual to the fixed area states only for $t\sim O(c)$ in the holographic CFTs. For $t\sim O(1)$ or $t\ll c\sim b$,  the one-point function of $T$ is still given by (\ref{onepointT}). It seems we could construct the geometry for these states. But the density of state $\mathcal{P}(t)$ no long scales as $e^{\sqrt{2 bt}}$, thus the EE $\log \mathcal{P}(t)$ in these states are not of $O(c)$. We don't expect they have well-defined bulk geometry.  
  
\section{Pure geometric states as superposition of fixed area states}
For the holographic CFT with large central charge $c$, our results in the above section show exactly that the state $|Phi\rangle_t$ can be explained as fixed area state if the $t$ is of the order of $c$. This gives us a new way to understand geometric states by decomposing them into fixed area states. To be more precise we have
\bea\label{decompositionvacuum}
|0\rangle=\sum_i e^{-\frac{b+t_i}{2}} |i\rangle |\bar i\rangle=\int_0^{\infty} dt\sqrt{\mathcal{P}(t)}e^{-\frac{b+t}{2}}|\Phi\rangle_t.  
\eea  
The reduced density matrix of $A$ is 
\bea\label{spectrumdecom}
\rho_A=\int_0^\infty dte^{-b-t}\mathcal{P}(t)\rho_{t,A}.
\eea
Actually, (\ref{spectrumdecom}) is just the spectrum decomposition of the operator $\rho_A$,   $P_t:=\mathcal{P}(t)\rho_{t,A}$ are projections into the Hilbert subspace with respect to the spectrum $e^{-b-t}$. The states $|\Phi\rangle_t$ are fixed area states if $t\sim O(c)$.  
However, the contributions  from $t\ll c$ are usually exponentially suppressed in the large $c$ limit. We can safely take the vacuum state of a holographic CFT as superposition of fixed area states by introducing a lower cut-off of the integral (\ref{decompositionvacuum}).\\
 For arbitrary pure geometric state $|\psi\rangle$ , the reduced density matrix of subsystem $A$ can be expressed as
 \bea\label{decompositionpsi}
 \rho^\psi_A=\int_0^{\infty}dt \mathcal{P}^\psi(t)e^{-b^\psi-t}\rho^\psi_{t,A}.
 \eea 
The density of eigenstates $\mathcal{P}^\psi(t)$ is given by
\bea\label{generaldensity}
&&\mathcal{P}^\psi(t)=\mathcal{L}^{-1}\left[ e^{n b+(1-n)S_n(\rho_{A}^{\psi})}\right](t)\nonumber \\
&&\phantom{\mathcal{P}^\psi(t)}=\frac{1}{2\pi i} \int_{\gamma_0-i\infty}^{\gamma_0+i\infty} dn e^{s_n},
\eea
with
\bea\label{sn}
s_n:=n(t+b)+(1-n)S_n(\rho_{A}^{\psi}),
\eea
where the $\gamma_0$ is chosen for the convergence of the integration, $S_n(\rho_{A}^{\psi})$ is the R\'enyi entropy of subsystem $A$ in the state $|\psi\rangle$. In general, it's hard to evaluate the R\'enyi entropy for aribrary states. For holographic theories, $S_n(\rho_{A}^{\psi})$ is expected to be of order $O(G^{-1})$ in the semiclassical limit $G\to 0$. For $t\sim O(G^{-1})$ we can evaluate the integral (\ref{generaldensity}) by saddle point approximation. That is to solve the equation 
\bea\label{equationkey}
\partial_n s_n=(t+b)-\partial_n[(1-n)S_n(\rho_{A}^{\psi})]=0.
\eea
In general, (\ref{equationkey}) is a complicated equation for $n$. Assume the solutions exist. 
If we have more than one solution, we should take the one that minimizes  $s_n$. With the solution $n^*=n^*(t)$ we have
\bea\label{stfunction}
s_{n^*}=\left[S_n(\rho_{A}^{\psi})+n(n-1)\partial_n S_n(\rho_{A}^{\psi})\right]_{n=n^*}.
\eea 
Using Dong's proposal of holographic R\'enyi entropy (\ref{holoRenyi}) we have
\bea\label{stholography}
s_{n^*}=\frac{\text{Area}(\mathcal{B}_{n^*})}{4G}.
\eea
Therefore, the density of eigenstates is given by
\bea\label{densitybranearea}
\mathcal{P}^\psi(t)\propto e^{\frac{\text{Area}(\mathcal{B}_{n^*})}{4G}}.
\eea
By definition the R\'enyi entropy of the state $\rho_{t,A}^\psi$ is independent with $n$, given by 
\bea\label{key1}
S_n(\rho_{t,A}^\psi)=\log \mathcal{P}^\psi(t)\simeq  \frac{\text{Area}(\mathcal{B}_{n^*})}{4G}.
\eea
Our results show the states $|\Psi\rangle_t^\psi$  have same property as the fixed area state. 
(\ref{equationkey}) and (\ref{key1}) give the dual relation between the parameter $t$ and the bulk fixed area, that is the area of the cosmic brane $\text{Area}(\mathcal{B}_{n})$.    
Supposed the geometry dual to $|\psi\rangle$ is $\mathcal{M}_\psi$. According to Dong's proposal of R\'enyi entropy the tension of the codimension-2 cosmic brane $\mathcal{B}_n$ is $\mu_n= \frac{n-1}{4nG}$. To obtain the geometry dual to the fixed area state $|\Phi\rangle^\psi_t$
one should insert a codimension-2 cosmic brane with tension $\mu_t=\frac{n^*-1}{4n^* G}$, where $n^*$ is the solution of the equation (\ref{equationkey}). If the equation has more than one solution,  we should take the one that minimizes the function $s_n$ (\ref{sn}).The cosmic brane backreacts on the geometry $\mathcal{M}_\psi$ and creats a conical defect with opening angle $ \theta:=2\pi \alpha_t=2\pi -8\pi G \mu_{n^*}$. The location of the cosmic brane coincides with the RT surface for subregion $A$ in the backreacted geometry. The role of the cosmic brane is like a sharp projection that maps the original geoemtry $\mathcal{M}_\psi$ to the fixed area geometry. The above results are consistent with the discussion in \cite{Dong:2018seb} by using the gravitational path integral.\\
As a check of the above statement, let's consider the vacuum state in AdS$_3$. Taking the R\'enyi entropy $S_{n}({\rho_A})=(1+\frac{1}{n})b$ into the equation (\ref{equationkey}),  we have the solution $n^*= \sqrt{b/t}$. The tension of the cosmic line is $\mu_t=\frac{1}{4G}(1-\sqrt{t/b})$ and the opening  angle of the conical defect line is $ \theta =2\pi \sqrt{t/b}$. The results are exactly consistent with our direct calculations in last section. \\
By using the expression of $\mathcal{P}^\psi(t)$ , arbitrary pure geometric state $|\psi\rangle$ can be seen as superposition of a series of the fixed area states,
\bea\label{decomposition}
|\psi\rangle =\int_0^\infty dt \sqrt{p^\psi_t} |\Phi\rangle^\psi_t,
\eea
where $p^\psi_t:=e^{\frac{\text{Area}{\mathcal{B}_{n^*}}}{4G}-b^\psi-t}$. Like the vacuum case we expect the contributions from small $t$ ($t\ll c$) of the above integration are  negligible. 
\section{Probability of the fixed area states}
The quantum error correction code interpretation of AdS/CFT suggests the coefficents $p^\psi_t$ of (\ref{decomposition}) can be associated with the on-shell action $I^\psi_t$ of the corresponding fixed area states $|\Phi\rangle^\psi_t$\cite{Dong:2018seb}\cite{Marolf:2020vsi}. The expected relation is $p^\psi_t=e^{-I^\psi_t}$. Using the result (\ref{decomposition}),  we have
\bea\label{actiongeneral}
I^\psi_t=b_\psi+t-\frac{\text{Area} \mathcal{B}_{n^*}}{4G}.
\eea
which depends on  the parmerter $t$. $p^\psi_t$ can be explained as the probability for the geometric state $|\psi\rangle$ to be  the fixed area state $|\Phi\rangle_t^\psi$. \\
For the vacuum case $|0\rangle$, $b_\psi=b$ and $\frac{\text{Area} \mathcal{B}_{n^*}}{4G}=2\sqrt{b t}$, the action $I_t=b(1-\sqrt{\frac{t}{b}})^2=b(1-\alpha)^2$, which is consistent with (\ref{ncopyaction}) and the results in \cite{Marolf:2020vsi}. The probability distribution $p_t:=e^{-I_t}$ has maximal value at $t=b$. In the semiclassical limit $G\to 0$ or $c\to \infty$, the distribution will approach to a delta function
$\delta(t-b)$. Therefore, $\rho_A$ can be approximated by the fixed area state $\rho_{t=b}$. One could check the EE of $\rho_A$ is same as $\rho_{t=b}$ in the leading order of $c$. Taking $t=b$ into (\ref{geometrydual}) we get same geometry as the vacuum AdS$_3$. However, we could find other probes that could distinguish the two states, see more discussions in \cite{Guo:2020roc}. This means the superposition among the fixed area states is important to understand the full properties of the  geometry dual to $|\psi\rangle$.    \\
We can also consider the unormalized $n$-copy state 
\bea
(\rho_A) ^n=\int_0^\infty dt p_t^n (\rho_t)^n\simeq \int_0^\infty dt \sqrt{\frac{b}{t}}e^{-n(b+t)+2\sqrt{b t}} \rho_{t,A}.\nonumber \\
\eea
It can be shown $(\rho_A) ^n\simeq e^{-(n-\frac{1}{n})b}\rho_{t=\frac{b}{n^2}}$.  This means the geometry of $n$-copy state is approximated by the fixed area state with $t=\frac{b}{n^2}$, which is the spacetime  inserting a cosmic brane with  tenson $\frac{n-1}{4G n}$. It is a consistent check with Dong's proposal of holographic R\'enyi entropy.
~\\
In general, $I^\psi_t$ is propotional to $1/G$. In the semiclassical limit $G\to 0$, we expect the probability $p_t^\psi$ has maximal value at $\bar t$, which is fixed by the equation $\partial_t I_t^\psi|_{t=\bar t}=0 $. It is not easy to find $\bar t$ by solving (\ref{actiongeneral}) and (\ref{equationkey}). Motivated by the vacuum case, we can fix $\bar t$ by requiring the EE of $\rho^\psi_{t=\bar t}$ is equal to the EE of $\rho^\psi_A$. This leads to $n^*(\bar t)=1$. Using (\ref{equationkey}) we find $\bar t=S(\rho^\psi_A)-b^\psi$.In \cite{Guo:2020roc} we show the one-point functions of local operators $\mathcal{O}$ in states $\rho^\psi_A$  are equal to the ones in $\rho^\psi_{t=\bar t}$ in the semiclassical limit $c\to \infty$. This leads to the result 
\bea\label{semilimit}
\int_0^\infty dt p^\psi_t \to \int_0^\infty dt \delta(t-\bar t)
\eea 
in the semiclassical limit $c\to \infty$ or $G\to 0$.

\section{Geometry and area operator}
The fixed area states $|\Phi\rangle^\psi_t$ can be taken as the basis of a given pure geometric state $|\psi\rangle$. 
We may introduce an operator $\hat{A}^\psi$, which is expected to satify the following conditions:
\begin{enumerate}[1)]
\item Positive semidefinite Hermitian and state-dependent opertor \cite{Papadodimas:2015jra}.
\item Fixed area states are its eigenstates.
\item Located in subsystem $A$ or  $\bar A$
\item Its expectation value in geometric state $|\psi\rangle$ divided by $4G$ gives the RT formula \cite{BottaCantcheff:2014qdq} and its fluctuation in $|\psi\rangle$ is supressed in the semiclassical limit $G\to 0$.
\end{enumerate}
The area operator $\hat{A}^\psi$ can be constructed by spectrum decomposition. The modular Hamiltonian $H^\psi_A$ has the spectrum decompostion as $H^\psi_A=\int_0^\infty dt (t+b^\psi) P^\psi_t$, where $P^\psi_t:= \mathcal{P}^\psi(t) \rho^\psi_{t,A}$. According to the operator theory  \cite{operatortheory}, we can define the new operators
\bea
F(H^\psi_A):=\int_0^\infty dt F(t+b^\psi) P_t^\psi,
\eea
where $F(x)$ is the functions of $x$\cite{com1}. The operators satisfy $F(H_A^\psi) |\Phi\rangle_t^\psi=F(t+b^\psi)|\Phi\rangle^\psi_A$. The area operator can be defined as 
\bea
\hat{A}^\psi =s(H_A^\psi-b^\psi)=\int_0^\infty dt s(t)P^\psi_t,
\eea
where $s(t):=\frac{6}{c}s_{n^*}$, $s_{n^*}$ is given by (\ref{stfunction}). If we further use the holographic proposal of R\'enyi entropy, the area operator is 
\bea
\hat{A}^\psi =\int_0^\infty dt \text{Area}(\mathcal{B}_{n^*})P^\psi_t,
\eea
where we used (\ref{stholography}) and the Brown-Henneaux relation $c=\frac{3}{2G}$. 
It is obvious that $\hat{A}^\psi |\Phi\rangle^\psi_t= \text{Area}(\mathcal{B}_{n^*})|\Phi\rangle^\psi_t$, $\text{Area}(\mathcal{B}_{n^*})$ is the area of the bulk RT surface for the geometry dual to the fixed area state $|\Phi\rangle_t^\psi$. 
The expectation value of $\hat A$ in  $|\psi\rangle$ is 
\bea
\langle \hat A^\psi \rangle_\psi= \int_0^\infty dt p^\psi_t \text{Area}(\mathcal{B}_{n^*}) =\int_0^\infty dt e^{-I^\psi_t}\text{Area}(\mathcal{B}_{n^*}).\nonumber
\eea
According to (\ref{semilimit}), we have 
\bea\label{RTquantum}
\langle \hat A^\psi \rangle_\psi\to \int_0^\infty dt \delta(t-\bar t)\text{Area}(\mathcal{B}_{n^*})= \text{Area}(\mathcal{B}_1),
\eea 
in the semiclassical limit $G\to 0$. $\text{Area}(\mathcal{B}_1)$ is just the area of the RT surface in the geometry dual to $|\psi\rangle$.  The RT formula can be expressed by area operator as
\bea
S(\rho^\psi_A)=\frac{\langle \hat A^\psi\rangle_\psi}{4G}.
\eea
By using the definition of the EE $S(\rho^\psi_A)=-tr (\rho^\psi_A \log \rho^\psi_A)=\langle \psi| H_A |\psi\rangle$, we have a nice result
\bea
&&\langle \psi| (H^\psi_A-\frac{\hat A^\psi}{4G}|\psi\rangle=\int_0^\infty dt e^{-I^\psi_t} ( t+b^\psi-\frac{\text{Area}(\mathcal{B}_n)}{4G}),\nonumber \\
&& \phantom{\langle \psi| (H^\psi_A-\frac{\hat A^\psi)}{4G}|\psi\rangle}\to (\bar t+b^\psi-\frac{\text{Area}(\mathcal{B}_1)}{4G})=0,
\eea
in the limit $G\to 0$. This can seen as the bulk dual of the modular Hamiltonian to the leading order in the $1/G$ expansion\cite{Jafferis:2015del}.  \\
To characterize the fluctuation of the area operator in the state $|\psi\rangle$ , we can define the uncertainty of the area operator $\langle \Delta \hat A^\psi\rangle_\psi:=\sqrt{\langle \hat (A^\psi)^2\rangle_\psi-\langle \hat A^\psi\rangle_\psi^2}$. By using (\ref{semilimit}), we can show $\langle \Delta \hat A^\psi\rangle_\psi=0$ in the limit $G\to 0$.\\
We show the results for vacuum state in Appendix C.
\section{Discussion}
In this paper we only focus on the pure geometric state. Some important modifications are necessary to generalize the results to the mixed states. \\
In the last section we only consider the leading order result in the expansion of  gravitational coupling $G$. The RT formula would receive correction at higher orders in $G$\cite{Faulkner:2013ana}. That would be interesting to consider the higher orders corrections, which is important to understand the quantum nature of spacetime.\\
Our constructed area operator  is expressed as superposition of projectors in CFTs.  It may be possible to find its bulk dual by reconstruction of the bulk operators in entanglement wedge\cite{Dong:2016eik}\cite{Faulkner:2017vdd}. \\
In the last part of the paper, we show the uncertainty of the area operator is vanishing to the leading order in $G$. This is the expected feature for geometric state, for which the quantum fluctuation should be suppressed. This property is similar to the constraints of geometric states\cite{Guo:2018fnv}, which are expressed as conditions for connected correlation functions of stress energy tensor. It would be interesting to find their possible connections.  \\

{\bf Acknowledgements} I am supposed by the National Natural
Science Foundation of China under Grant No.12005070 and the Fundamental Research
Funds for the Central Universities under Grants NO.2020kfyXJJS041

\appendix
\section{Appendix A: geometry of the fixed area states}
The AdS$_3$ metric in the Poincar\'e coordinate is 
\bea
ds^2=\frac{d\eta^2+d\xi d\bar \xi}{\eta^2},
\eea
where $\xi,\bar \xi$ are the complex coordinates. The dual CFT lives on the boundary $\eta=0$.  The bulk coordinate transformation  associated with a boundary conformal mapping $\xi=f(z)$ and $\bar \xi =\bar f(\bar z)$ is given by 
\bea\label{bulkcoordinatetrans}
&&\xi=f(z)-\frac{2y^2 f'(z)^2 \bar f''(\bar z)}{4f'(z)\bar f'(\bar z)+u^2f''(z)\bar f''(\bar z)},\nonumber \\
&&\bar \xi= \bar f(\bar z)-\frac{2y^2 \bar f'(\bar z)^2  f''( z)}{4f'(z)\bar f'(\bar z)+u^2f''(z)\bar f''(\bar z)},\nonumber\\
&&\eta=\frac{4u(f'(z)\bar f'(\bar z))^{3/2}}{4f'(z)\bar f'(\bar z)+u^2f''(z)\bar f''(\bar z)}.
\eea 
The bulk metric in the coordinate $(u,z,\bar z)$ is 
\bea\label{smmetric}
&&ds^2=\frac{du^2}{u^2}+\frac{L(z)}{2}dz^2+\frac{\bar L(\bar z)}{2}d\bar z^2\nonumber \\
&&\phantom{ds^2}+\left(\frac{1}{u^2}+\frac{u^2}{4}L(z) \bar L(\bar z)\right)dz d\bar z,
\eea
with
\bea
&&L(z)=\{ f(z);z\}:=\frac{3f''(z)^2-2f'(z)f'''(z)}{2f'(z)^2} ,\nonumber \\
&&\bar L(\bar z)=\{ \bar f(\bar z);\bar z\}:=\frac{3\bar f''(\bar z)^2-2\bar f'(\bar z)\bar f'''(\bar z)}{2\bar f'(\bar z)^2},
\eea 
where $\{f(z);z\}$ is the Schwarzian derivative. Using the transformation law of the stress energy tensor of the boundary theory, we have $L(z)=-\frac{12}{c}tr(\rho T(z))$ and $\bar L(\bar z)=-\frac{12}{c}tr(\rho \bar T(\bar z))$, where $\rho$ is the state of the boundary CFT. \\
We can use the above coordinate transformation associated the conformal mapping $\xi\to w$ to evaluate the geodesic line homology to $A$.  By the conformal mapping $\xi=g(w):=\left(\frac{R+w}{R-w}\right)^\alpha$,  the ending point of $A$ $w_2=R$ is mapped to $\xi=\infty$. To regularize the coordinate in the $\xi$-plane we choose the ending points of $A$ as $w_1=-R+\epsilon$ and $w_2=R-\epsilon$, where $\epsilon$ is the UV cut-off.  The associated bulk coordinate transformation is given by (\ref{bulkcoordinatetrans}) with  replacing $u\to y$, $f(z)\to g(w)$ and $\bar g(\bar z)\to \bar g(\bar w)$. \\
By RT formula the EE of subsystem $A$ is related to the geodesic line $\gamma_A$ connecting $(y,w)=(\epsilon,w_1)$ and $(y,w)=(\epsilon,w_2)$. Their images in the Poincar\'e coordinate   are $(\eta,\xi)= (\eta_1, g(w_1))$ and $(\eta,\xi)= (\eta_2, g(w_2))$, where $\eta_{1(2)}$ is obtained by using the third equation of (\ref{bulkcoordinatetrans}) and taking $u\to \epsilon$, $f(z)\to g(w)|_{w=w_{1(2)}}$. The length geodesic line  connecting $(\eta_1, g(w_1))$ and $(\eta_2, g(w_2))$ is 
\bea
L_{\gamma_A}=\log \frac{(\xi_1-\xi_2)(\bar \xi_1-\bar \xi_2)}{\eta_1\eta_2}.
\eea
With some calculations we have
\bea
S_A=\frac{L_{\gamma_A}}{4G}=\frac{\alpha}{2G}\log \frac{2R}{\epsilon}=\alpha \frac{c}{3}\log \frac{2R}{\epsilon},
\eea
where we use the Brown-Henneaux relation $c=\frac{3}{2G}$.\\
In the main text we use the condition $S_n(\rho_{t,A})=S(\rho_{t,A})$ to get $\gamma=\gamma_A$. We could obtain $\gamma=\gamma_A$ more diretly on the z-plane with the coordiante transformation $z=\frac{R-w}{R+w}$.This is a global conformal transformation which maps the interval $A$ to the half line $[0,+\infty)$. We can work out the corresponding bulk metric
\bea\label{bulkmetricz}
&&ds^2=\frac{du^2}{u^2}+\frac{\alpha^2-1}{4z^2}dz^2+\frac{\alpha^2-1}{4\bar z^2}d\bar z^2\nonumber \\
&&\phantom{ds^2}+\left(\frac{1}{u^2}+\frac{(\alpha^2-1)^2}{16}u^2\right)dz d\bar z,
\eea 
where $u$ is the holographic coordinate. With a further conformal map $\xi=z^{-\alpha}$ we arrive at the $\xi$-plane. Actually, the transformation $\xi=\left(\frac{R+w}{R-w}\right)^\alpha$ is given by the combined two  maps $w\to z \to \xi$. For the conformal mapping $\xi=\left(\frac{R+w}{R-w}\right)^\alpha, \bar \xi =\left(\frac{R+\bar w}{R-\bar w}\right)^\alpha$, the bulk metric transforms to (\ref{geometrydual}). For the conformal mapping $\xi=z^{-\alpha}$, $\bar \xi= \bar z^{-\alpha}$, we get the bulk metric (\ref{bulkmetricz}). With a conformal mapping $z=\frac{R+w}{R-w}$ the bulk metric (\ref{geometrydual}) transforms to (\ref{bulkmetricz}).\\
By symmetry the conical defect line $\gamma$ should be the $u$-axis, which is same as the geodesic line $\gamma_A$.\\

\section{Appendix B: Action for $\mathcal{M}_n$}
To evaluate R\'enyi entropy we need to calculate the on-shell action of $\mathcal{M}_n$. We will focus on 3D.  The action consists of three parts,
\bea
I_{g}=I_{EH}+I_{GH}+I_{ct},
\eea
with
\bea
&&I_{EH}=-\frac{1}{16\pi G} \int d^3x \sqrt{g}(R+2),\nonumber \\
&&I_{GH}=-\frac{1}{8\pi G} \int_{bdy} \sqrt{h} K\nonumber \\
&&I_{ct}=\frac{1}{16\pi G} \int_{bdy} \sqrt{h},
\eea
where ``byd'' means the boundary of the bulk. The general metric of AdS$_3$ is given by (\ref{smmetric}). The boundary is taken to be the surface $u=\epsilon$. 
The term $I_{EH}$ involves the integration over the bulk, we should take an IR cut-off $u=u_{IR}$ to regularize it. We will follow the strategy of \cite{Hung:2011nu} to fix $u_{IR}$ by the condition $det(g)=0$, which leads to the solution 
\bea
u_{IR}=\frac{\sqrt{2}}{(L(z)\bar L(\bar z))^{1/4}}.
\eea
With some calculations we have
\bea
&&I_{EH}=-\int dzd\bar z \left[-\frac{1}{16\pi G\epsilon^2} +\frac{\sqrt{L(z)\bar L (\bar z)}}{16\pi G}\right],\nonumber \\
&&I_{GH}=-\int dzd\bar z \left[\frac{1}{8\pi G\epsilon}\right],I_{ct}=\int dzd\bar z \frac{1}{16\pi G\epsilon}.
\eea
The total action is 
\bea
I_{g}=-\frac{1}{16\pi G}\int dzd\bar z\sqrt{L(z)\bar L (\bar z)}.
\eea
To construct the fixed area state the cosmic brane is necessary. The cosmic brane backreacts the original geometry, thus the one-point function of stress energy tessor $T_{\mu\nu}$ should depends on the tension of cosmic brane.  The action also contains the contributions from the cosmic brane, which is related to the opening angle of the conical defect and the area of the cosmic brane. 
For metric with conical defect the Ricci scalar contains a delta function.A new term associated with the length of the cosmic line is 
\bea
I_{brane}=-\mu \int_{\gamma }d l, 
\eea
where $\mu$ is the tension of the cosmic brane.\\
The above results can be used to calculate the on-shell gravity action of the fixed area states.  Consider the interval $A=[-R,R]$ on the $w$-plane. The geometry is given by the metric (\ref{geometrydual})  The on-shell action of the geometry is given by 
\bea\label{actionM1}
&&I_g(\mathcal{M}_1)=-\frac{1}{16\pi G} \int dwd\bar w \sqrt{L_t \bar L_t},\\
&&\phantom{I_g(\mathcal{M}_1)}=-\frac{1}{16\pi G} (1-\frac{t}{b})\int dw d\bar w \frac{2R^2}{|R-w|^2|R-\bar w|^2}.\nonumber \\
~
\eea  
The above integration has singularity at the ending point of $A$. The integration of (\ref{actionM1}) is evaluated in the $w$-plane with cut-off $|w-R|\ge \epsilon$ and $|w+R|\ge \epsilon$.  The same integration is done in section.3.2 in \cite{Hung:2011nu}. The integral can be reduced to two integrated terms around the cut-off circle around the ending point of $A$. The result is 
\bea\label{intgeo}
&&I_g(\mathcal{M}_1)=-\frac{1}{64\pi G}\left( 1-\frac{t}{b} \right) \left[\oint_{w=-R}+\oint_{w=R}\right]d\theta \epsilon  (\chi \partial_r \chi)\nonumber \\
&&\phantom{I_g(\mathcal{M}_1)}=\frac{1}{4G}\left( 1-\frac{t}{b} \right) \log\frac{2R}{\epsilon}=(1-\alpha^2)b,
\eea
where $\chi:=2\log |\frac{w+R}{w-R}|$, $\alpha=\sqrt{\frac{t}{b}}$ and $b=\frac{1}{4G}\log\frac{2R}{\epsilon}$.\\
The geometry $\mathcal{M}_1$ has conical defect at the line $\gamma$. The tension of the cosmic brane is $\mu=\frac{1-\alpha}{4G}$. The action from the cosmic line is
\bea
I_{brane}=\frac{1-\alpha}{4G}L_\gamma,
\eea
where $L_\gamma$ is the length of the cosmic line. \\
For the fixed area states the action of $\mathcal{M}_n$ is generally expeceted to be 
\bea\label{actiongeneral}
I(\mathcal{M}_n)=n I_{g}(\mathcal{M}_1)+\frac{n \Delta \theta_n -2\pi}{8\pi G}A,
\eea
where $I_g$ denotes the contributions from the geometry, $A$ is the area of the cosmic brane. The second term of the right hand side of (\ref{actiongeneral}) is the contribution from cosmic brane. The ansatz of the action of $\mathcal{M}$ leads to the result that R\'enyi entropy in independent with $n$. \\
For the n-cpoy spacetime $\mathcal{M}_n$, the action from the geomtry is just $n$ times $I_g(\mathcal{M}_1)$ since the range of $\theta$ in  (\ref{intgeo}) is $2n\pi$. The tension of the cosmic line is $\mu_n=\frac{1-n\alpha}{4G}$. The total action is 
\bea
I_{tot}(\mathcal{M}_n) =n I_{g}(\mathcal{M}_1)+\frac{n\alpha-1}{4G} L_\gamma,
\eea
 which is consistent with the ansatz (\ref{actiongeneral}).
\section{Appendix C: Area operator for vacuum state}

We can study more details of the area operator for the vacuum case $|\psi\rangle=|0\rangle$. The area operator is given by
\bea
\hat A=4G\int_0^\infty dt  (2\sqrt{b t}) P_t,
\eea
where we have used the fact $\text{Area}(\mathcal{B})_{n^*}\simeq 2\sqrt{bt}$.
The expectation value of $\hat A$ in vacuum is
\bea
&&\langle \hat A \rangle= \int_0^\infty dt \mathcal{P}(t)e^{-b-t} (2\sqrt{b t}),\nonumber \\
&&\phantom{\langle \hat A \rangle}=\sqrt{\pi} b^{3/2}e^{-b/2}\left[I_0(\frac{b}{2})+I_1(\frac{b}{2})\right]\simeq 8 Gb.
\eea
In the last step we use $I_n(x)\simeq \frac{e^x}{\sqrt{2\pi x}}$ for large $x$. The result is consistent with the classical RT formula $S_A=\frac{\langle \hat A\rangle}{4G}=2b$.\\
The expectation value of $\hat A^2$ is
\bea
&&\langle \hat A^2 \rangle= (4G)^2 \int_0^\infty dt \mathcal{P}(t)e^{-b-t} (2\sqrt{b t})^2 \simeq 64(Gb)^2.
\eea
Thus the uncertainty of the operator $\hat A$ is vanishing to the leading order in $G$,
\bea
\langle\Delta \hat A\rangle=\sqrt{\langle \hat A^2\rangle-\langle \hat A\rangle^2}=0.
\eea
Motivated by the moments methods of a random variable, we can further check the third moment of the area operator defined as
\bea
&&\langle\Delta \hat A\rangle_3:= \langle (\hat A-\langle \hat A\rangle)^3\rangle\nonumber \\
&&= \langle \hat A^3\rangle-2\langle \hat A^2\rangle \langle \hat A\rangle+2\langle \hat A\rangle \langle \hat A\rangle^2-\langle \hat A\rangle^3.
\eea
With some calculations we find 
\bea
\langle \Delta \hat A\rangle_3= 512 (Gb)^2 G.
\eea
The above result is of order $O(G)$. In the semiclassical limit $G\to 0$, we have $\langle \Delta \hat A\rangle_3\to 0$. One could show the $n$-th moments $\langle\Delta \hat A\rangle_n:= \langle (\hat A-\langle \hat A\rangle)^n\rangle$ are vanishing in the limit $G\to 0$.

\begin{thebibliography}{00}
\bibitem{com1}
The reduced density matrix $\rho_A$ is given by the function $F(x)=e^{-x}$.
\end{thebibliography}
\end{document}